\documentclass{article}
\usepackage{amsmath,graphics,epsfig}

\def\openone{\leavevmode\hbox{\small1\kern-3.3pt\normalsize1}}

\allowdisplaybreaks
\begin{document}

\title{On Modelling adiabatic $N$-soliton interactions in
external potentials }

\author{V.\ S.\ Gerdjikov$^1$, B.\ B.\ Baizakov$^2$
        and M.\ Salerno$^2$ \\
\\
\normalsize{\it $^1$ Institute for Nuclear Research and Nuclear Energy,
                          1784 Sofia, Bulgaria} \\
\normalsize{\it $^2$ Department of Physics "E.R.Caianiello" and Istituto
Nazionale} \\
\normalsize{\it di Fisica della Materia, University of Salerno,
I-84081 Baronissi, Italy}}
\date{}

\maketitle

\begin{abstract}
We analyze perturbed version of the complex Toda chain (CTC) in an
attempt to describe the adiabatic  $N$-soliton train interactions of the
perturbed nonlinear Schr\"odinger equation (NLS). We study perturbations
with weak quadratic and periodic external potentials by both analytical
and numerical means.  The perturbed CTC adequately models the $N $-soliton
train dynamics for both types of potentials. As an application of the
developed theory we consider the dynamics of a train of matter - wave
solitons confined in a parabolic trap and optical lattice, as well as
tilted periodic potentials.

\end{abstract}


\section{Introduction}

The $N $-soliton train interactions for the nonlinear
Schr\"odinger equation (NLS) and its perturbed versions
\begin{equation}\label{eq:nls}
iu_t + {1  \over 2 }u_{xx} + |u|^2u(x,t)=i\epsilon R[u],
\end{equation}
started with the pioneering paper \cite{Karp}, by now has been extensively
studied (see \cite{GKUE,GUED,Arnold,GKDU,Liss} and references therein).
Several other nonlinear evolution equations (NLEE) were also studied,
among them the modified NLS equation \cite{KodHas,IG_BJ,Val&I,GDY,GU},
some higher NLS equations \cite{Liss}, the Ablowitz-Ladik system
\cite{Dokt} and others.

Below we concentrate on the perturbed NLS eq. (\ref{eq:nls}).
By $N $-soliton train we mean a solution of the (perturbed) NLS fixed up
by the initial condition:
\begin{eqnarray}\label{eq:IC}
u(x,t=0) &=& \sum_{k=1}^{N} u_k^{\rm 1s}(x,t=0) ,\qquad
u_k^{\rm 1s}(x,t) = {2\nu _k e^{i\phi _k}\over \cosh z_k }, \\
\label{eq:nls-1s}
z_k(x,t) &=& 2\nu _k (x-\xi _k (t)), \qquad \xi_k(t) = 2\mu _kt +
\xi_{k,0} \\
\phi _k(x,t)&=& {\mu _k \over \nu _k} z_k + \delta _{k}(t),\qquad
\delta _k(t) = W_kt +\delta _{k,0},
\end{eqnarray}
Each soliton has four parameters: amplitude $\nu_k$, velocity $\mu_k$,
center of mass position $\xi_k$ and phase $\delta _k$.
The adiabatic approximation uses as a small parameter $\varepsilon_0
\ll 1$ the soliton overlap which falls off exponentially with the distance
between the solitons. Then the soliton parameters must satisfy
\cite{Karp}:
\begin{eqnarray}\label{eq:ad-ap}
&& |\nu _k-\nu _0| \ll \nu _0, \quad |\mu _k-\mu _0| \ll \mu _0,
\quad |\nu _k-\nu _0| |\xi_{k+1,0}-\xi_{k,0}| \gg 1,
\end{eqnarray}
where $\nu _0 = {1  \over N }\sum_{k=1}^{N}\nu _k$, and $ \mu _0 =
{1 \over N }\sum_{k=1}^{N}\mu _k$ are the average amplitude and
velocity respectively. In fact we have two different scales:
\[ |\nu _k-\nu _0| \simeq \varepsilon_0^{1/2}, \qquad
|\mu _k-\mu _0| \simeq \varepsilon_0^{1/2}, \qquad
|\xi_{k+1,0}-\xi_{k,0}| \simeq \varepsilon_0^{-1}.
\]
One can expect that the approximation holds only for such times
$t$ for which the set of $4N$ parameters of the soliton train
satisfy (\ref{eq:ad-ap}).

Equation (\ref{eq:nls}) finds a number of applications
to nonlinear optics and for $R[u]\equiv 0 $ is integrable
via the inverse scattering transform method \cite{ZMNP,FaTa}.
The $N $-soliton train
dynamics in the adiabatic approximation is modelled by a complex
generalization of the Toda chain \cite{MFM}:
\begin{equation}\label{eq:ctc}
\frac{{\rm d}^2Q_j}{{\rm  d}t^2}= 16\nu _0^2\left( e^{Q_{j+1}-Q_j}-
e^{Q_j-Q_{j-1}} \right) , \qquad j=1,\dots ,N.
\end{equation}
The complex-valued $Q_k $ are expressed through the soliton
parameters by:
\begin{equation}\label{hnls-Q}
Q_{k}(t) = 2i\lambda  _0 \xi_k(t) + 2k\ln (2\nu _0 ) +
i(k\pi -\delta_{k}(t)  -\delta_0 ),
\end{equation}
where $\delta _0=1/N \sum_{k=1}^{N}\delta _k $ and $\lambda _0=\mu _0+i\nu
_0 $. Besides we assume free-ends conditions, i.e., $e^{-Q_0}\equiv
e^{Q_{N+1}}\equiv 0$.

Note that the $N $-soliton train is {\em not} an $N $-soliton solution.
The spectral data of the corresponding Lax operator $L $ is nontrivial
also on the continuous spectrum of $L $. Therefore the analytical results
from the soliton theory can not be applied. Besides we want to treat
solitons moving with equal velocities and also to account for the effects
of possible nonintegrable perturbations $R[u] $.

The present paper extends the results of several previous ones:  see Refs.
\cite{GKUE,GUED,Arnold,GKDU,GU,G_Gal02}.  Recently with the realization of
Bose-Einstein condensation of dilute atomic gases it became important to
study NLS equation with additional potential term $iR[u]=V(x)u(x,t) $, see
\cite{BCCD,carretero}. We continue the analysis in \cite{Liss} and in
our more recent reports \cite{GBS}, of the corresponding perturbed CTC
(PCTC) model for quadratic and periodic potentials $V(x) $. Our results
give additional confirmation of the stabilization properties of the
periodic potentials observed in \cite{WB,UGL} in a different physical
setup.

We also pay attention to the so-called tilted periodic potentials,
which are superpositions of periodic and linear potentials. The effect of
the linear potential is that it can, if it is strong enough, overcome the
confining effect of the periodic potential. As a result we can have
one or more of the solitons extracted out of the train. We also
demonstrate that the PCTC provides an adequate description also for these
types of potentials.

\section{The importance of the CTC model}

The fact \cite{MFM,Moser} that the CTC, like the (real) Toda chain (RTC),
is a completely integrable Hamiltonian system allows one to analyze
analytically the asymptotic behavior of the $N$-soliton trains.
However unlike the RTC, the CTC has richer variety of dynamical
regimes \cite{GKUE,GKDU,GEI} such as:
\begin{itemize}
\item asymptotically free motion if $v_j\neq v_k $ for $j\neq k $; this is
the only dynamical regime possible for RTC;

\item $N $-s bound state if $v_1 =\dots = v_N $ but
  $\zeta _k\neq \zeta _j $ for $k\neq j $;

\item  various intermediate (mixed) regimes; e.g., if $v_1= v_2 >
\dots > v_N $ but $\zeta _k\neq \zeta _j $ for $k\neq j $ then
we will have a bound state of the first two solitons while all the
others will be asymptotically free;

\item singular and degenerate regimes if two or more of the eigenvalues
of $L$ become equal, e.g., $\zeta _1=\zeta _2 \dots $ and $\zeta _j\neq
\zeta _k $ for $2<j\neq k $.
\end{itemize}

By $\zeta_k=v_k+iw_k$ above we have denoted the eigenvalues of the Lax
matrix $L$ in the Lax representation $L_\tau = [M,L]  $ of the CTC where:
\begin{eqnarray}\label{eq:L-M}
L&=& \sum_{k=1}^{N} b_k E_{kk} + \sum_{k=1}^{N-1} a_k ( E_{k,k+1} +
E_{k+1,k}), \\
b_k &\equiv & -{1 \over 2} {dQ_k  \over d\tau } = {1\over 2 }
(\mu _k+i\nu_k) , \qquad a_k = {1\over 2 } \exp((Q_{k+1}-Q_k)/2).
\nonumber
\end{eqnarray}
and the matrices $E_{kp} $ are defined by
$(E_{kp})_{ij}=\delta_{ki}\delta_{pj}$.  The eigenvalues $\zeta _k $ of $L
$ are time independent and complex-valued along with the first components
$\eta _k=\vec{z}_1^{(k)}$ of the normalized eigenvectors of $L$:
\begin{equation}\label{eq:L-v}
L\vec{z}^{(k)} = \zeta _k \vec{z}^{(k)}, \qquad
(\vec{z}^{(k)},\vec{z}^{(m)})=\delta _{km}.
\end{equation}
The set of $\{\zeta_k = v_k +i w_k, \; \eta_k = \sigma_k +
i\theta_k\} $ may be viewed as the set of action-angle variables
of the CTC.

Using the CTC model  one can determine the asymptotic regime of
the $N $-soliton train.  Given the initial parameters  $\mu _k(0),
\nu _k(0), \xi_{k}(0), \delta _k(0) $ of the $N $-soliton train
one can calculate the matrix elements $b_k$ and $a_k$ of $L$ at
$t=0$. Then solving  the characteristic equation on $L|_{t=0}$ one
can calculate the eigenvalues $\zeta_k$ to  determine the
asymptotic regime of the $N $-soliton train \cite{GKUE,GKDU}.
Another option is to impose on $\zeta_k$ a specific constraint,
e.g. that all $\zeta _k $ be purely imaginary, i.e. all $v_k=0$.
This will provide a set of algebraic conditions  $L|_{t=0}$, and
on the initial soliton parameters $\mu _k(0), \nu _k(0),
\xi_{k}(0), \delta _k(0) $ which  characterize the region in the
soliton parameter space responsible for the $N$-soliton bound
states.

\section{The perturbed NLS and perturbed CTC}
\label{sec:Pert}

We will consider several specific choices $R^{(p)}[u] $ of
perturbations, $p=1,2,\dots $ in (\ref{eq:nls}). In the adiabatic
approximation the dynamics of the soliton parameters can be
determined by the system (see \cite{Karp} for $N=2 $ and
\cite{GKUE,GKDU} for $N>2 $):
\begin{eqnarray}\label{eq:nuk0}
{d\lambda _k \over dt} &=& -4\nu _0 \left(e^{Q_{k+1}-Q_k} -
e^{Q_k -Q_{k-1}} \right) +  M^{(p)}_k +i N^{(p)}_k , \\
\label{eq:xik0}
{d \xi_k \over dt} &=& 2 \mu_k +   \Xi^{(p)}_k  ,\qquad
{d \delta_k \over dt} = 2 (\mu_k^2 + \nu_k^2)+   X_k^{(p)} ,
\end{eqnarray}
where $\lambda _k=\mu _k+i\nu _k $ and $X_k^{(p)}  = 2 \mu_k \Xi_k^{(p)}
+ D_k^{(p)} $.  The right hand sides of Eqs.
(\ref{eq:nuk0})--(\ref{eq:xik0}) are determined by $R_k^{(p)}[u]$ through:
\begin{eqnarray}\label{eq:Nk0}
N_k^{(p)}  &=& {1 \over 2} \int_{-\infty}^{\infty} {dz_k \over \cosh z_k
}\, \mbox{Re}\,  \left( R_k^{(p)} [u] e^{-i\phi_k} \right) ,\\
\label{eq:Mk0}
M_k^{(p)}  &=& {1 \over 2} \int_{-\infty}^{\infty} {dz_k  \, \sinh z_k
\over \cosh^2 z_k }\,\mbox{Im}\,\left( R_k^{(p)} [u] e^{-i\phi_k} \right),
\\
\label{eq:Xik0}
\Xi_k^{(p)}  &=& {1 \over 4 \nu_k^2} \int_{-\infty}^{\infty} { dz_k \,
z_k\over \cosh z_k }\, \mbox{Re}\, \left( R_k^{(p)} [u] e^{-i\phi_k}
\right),\\
\label{eq:Dk0}
 D_k^{(p)}  &=& {1 \over 2 \nu_k} \int_{-\infty}^{\infty} {dz_k \,
( 1 - z_k \tanh z_k)  \over \cosh z_k }
\mbox{Im}\, \left( R_k^{(p)} [u] e^{-i\phi_k} \right) .
\end{eqnarray}

Inserting (\ref{eq:nuk0}), (\ref{eq:xik0}) into (\ref{hnls-Q}) we derive:
\begin{eqnarray}
{dQ_k  \over dt } &=& -4\nu _0\lambda _k + {2k \over\nu _0 }
\mathcal{N}_{0}^{(p)} + 2i\xi_k   \left( \mathcal{M}_{0}^{(p)} +
i \mathcal{N}_{0}^{(p)}\right) \nonumber\\
\label{eq:pctc}
&+& i   \left( 2\lambda _0 \Xi_{k}^{(p)} - X_{k}^{(p)} -
\mathcal{X}_{0}^{(p)}\right),\\
\mathcal{N}_{0}^{(p)} &=& {1\over N} \sum_{j=1}^{N} N_{j}^{(p)}, \qquad
\mathcal{M}_{0}^{(p)} = {1\over N} \sum_{j=1}^{N} M_{j}^{(p)}, \qquad
\mathcal{X}_{0}^{(p)} = {1\over N} \sum_{j=1}^{N} X_{j}^{(p)}.
\nonumber
\end{eqnarray}

In deriving eq. (\ref{eq:pctc}) we have kept terms of the order
$\Delta \nu _k \simeq \mathcal{O}(\sqrt{\epsilon_0 }) $ and
neglected terms of the order $\mathcal{O}(\epsilon_0 ) $. The
perturbations result in that $\nu _0 $ and $\mu _0 $ may become
time-dependent. Indeed, from (\ref{eq:nuk0}) we get:
\begin{equation}\label{eq:lam0}
{d\mu _0 \over dt } = \mathcal{M}_{0}^{(p)} , \qquad
{d\nu _0 \over dt } = \mathcal{N}_{0}^{(p)} ,
\end{equation}

The small parameter $\epsilon _0 $ can be related to the initial
distance $r_0=|\xi_2 -\xi_1|_{t=0} $ between the two solitons. Assuming
$\nu _{1,2}\simeq \nu _0 $ we find:
\begin{equation}\label{eq:eps0}
\epsilon _0 = \int_{-\infty }^{\infty } dx \left| u_{1}^{\rm 1s}(x,0)
u_{2}^{\rm 1s}(x,0)\right| \simeq 8\nu _0 r_0e^{-2\nu _0r_0}.
\end{equation}
In particular, (\ref{eq:eps0}) means that $\epsilon _0\simeq 0.01 $ for
$r_0\simeq 8 $ and $\nu _0=1/2 $.

We assume that initially the solitons are ordered in such a way that
$\xi_{k+1}-\xi_k \simeq r_0$.  One can check \cite{GUED,GU} that
$N_k^{(p)}\simeq M_k^{(p)} \simeq \exp (- 2\nu _0|k-p| r_0) $. Therefore
the interaction terms between the $k $-th and $k\pm 1 $-st solitons will
be of the order of $e^{-2\nu _0r_0} $; the interactions between $k $-th
and $k\pm 2 $-nd soliton will of the order of $e^{-4\nu_0r_0} \ll
e^{-2\nu_0r_0}$.

The terms $\Xi_{k}^{(0)} $, $X_{k}^{(0)} $ are of the order of
$r_0^a\exp(-2\nu_0r_0) $, where $a=0 $ or $1 $. However they can be
neglected as compared to $\widetilde{\mu }_k $ and $\widetilde{\nu }_k $,
where
\begin{equation}\label{eq:mu_k}
\widetilde{\mu }_k = \mu _k-\mu _0 \simeq \sqrt{\epsilon_0 }, \qquad
\widetilde{\nu }_k = \nu _k-\nu _0\simeq \sqrt{\epsilon_0 },
\end{equation}

The corrections to $N_{k}^{(p)} $, \dots, coming from the terms
linear in $u $ depend only on the parameters of the $k $-th
soliton; i.e., they are `local' in $k $. The nonlinear terms in $u
$ present in $iR^{(p)}[u] $ produce also `non-local' in $k $ terms
in $N_{k}^{(p)} $, \dots.

\subsection{Nonlinear gain and second order dispersion }\label{ssec:6.1}

Consider the NLS eq. (\ref{eq:nls}) with
\begin{equation}\label{eq:Ru_1}
R[u] = c_0u + c_2 u_{xx} + d_0 |u|^2u,
\end{equation}
where $c_0 $, $c_2 $ and $d_0 $ are  real constants, see \cite{GU}.
Another important factor is the order of magnitude of the perturbation
coefficients $c_0 $, $c_2 $ and $d_0 $ in (\ref{eq:Ru_1}).
If we take them to be of the order of $\epsilon _0 $ we find that
the $N $-soliton train evolves according to:
\begin{equation}\label{eq:p-ctc}
\frac{d^2Q_k }{ dt^2 } = U_{00} + 16\nu _0^2 \left(
e^{Q_{k+1}-Q_{k}} - e^{Q_{k}-Q_{k-1}} \right),
\end{equation}
where for $\mu _0=0 $ we get $U_{00}=-{8i\nu _0^2 \over 3 }
\left( 3c_{0} +8\nu _0^2d_{0}- 4c_{2} \nu _0^2 \right) $.
This form of perturbed CTC (\ref{eq:p-ctc})  can be solved exactly:
\[
Q_k(t) = \frac{1 }{2} U_{00}t^2 + V_{00}t+ Q^{(0)}_k(t),
\]
where $Q^{(0)}_k(t) $ is a solution of the unperturbed CTC and
$V_{00} $ is an arbitrary constant. In this case the effect of
the perturbation will be an overall motion of the center of mass
of the $N$-soliton train. The relative motion of the solitons will
remain the same.
For larger values of the coefficients $c_0 $, $c_2 $ and $d_0 $,  e.g., of
the order of $\sqrt{\epsilon_0}$ the corresponding dynamical system
is more complicated and has to be treated separately.

\subsection{Quadratic and periodic potentials}\label{Quadratic}

Let $iR[u]=V(x)u(x,t) $. Our first choice for $V(x) $ is a quadratic one:
\begin{equation}\label{eq:P3.5}
V^{(1)}(x) = V_2x^2 + V_1x + V_0.
\end{equation}
Skipping the details we get the results:
\begin{subequations}\label{eq:P3.4}
\begin{eqnarray}\label{eq:P3.4a}
N_{k}^{\rm (1)}  &=& 0, \qquad
M_{k}^{\rm (1)}  = -V_2 \xi_k - {V_1  \over 2 } , \\
\label{eq:P3.4c}
\Xi_{k}^{\rm (1)}  &=& 0, \qquad
D_{k}^{\rm (1)}  = V_2\left( {\pi^2  \over 48\nu _k^2 } -
\xi_k^2\right) -V_1 \xi_k -V_0,
\end{eqnarray}
\end{subequations}
and $X_{k}^{\rm  (1)}  = D_{k}^{\rm (1)}  $. As a result the corresponding
PCTC takes the form \cite{Liss}:
\begin{eqnarray}\label{eq:P4.3a}
&&{d(\mu _k + i\nu _k ) \over dt } = -4\nu _0 \left( e^{Q_{k+1}-Q_{k}}
-  e^{Q_{k}-Q_{k-1}} \right) -V_2\xi_k - {V_1  \over 2 }, \\
\label{eq:P4.3b}
&&{dQ_k  \over dt } = - 4\nu _0(\mu _k+i\nu _k) - iD_{k}^{\rm (1)}
- {i \over N } \sum_{j=1}^{N} D_{j}^{\rm (1)} ,
\end{eqnarray}

If we now differentiate (\ref{eq:P4.3b}) and make use of (\ref{eq:P4.3a})
we get \cite{Liss}:
\begin{eqnarray}\label{eq:P4.4}
{d^2 Q_k  \over dt^2 } &=& 16\nu _0^2 \left( e^{Q_{k+1}-Q_{k}}
-  e^{Q_{k}-Q_{k-1}} \right) + 4\nu _0 \left(V_2\xi _k + {V_1  \over
2}\right) - i{d D_{k}^{\rm (1)}  \over dt }- { i \over N }
\sum_{j=1}^{N} {d D_{j}^{\rm (1)}  \over dt}. \nonumber
\end{eqnarray}
It is reasonable to assume that $V_2\simeq {\cal  O}(\epsilon_0/N )$; this
ensures the possibility to have the $N $-soliton train `inside' the
potential. It also means that both the exponential terms and the
correction terms $M_{k}^{\rm (1)}  $ are of the same order of magnitude.
From eqs. (\ref{eq:P4.3a}) and (\ref{eq:P4.3b}) there follows that $d\nu
_0/dt =0 $ and:
\begin{equation}\label{eq:mu0-xi0}
{d\mu _0 \over dt } = -V_2 \xi_0 -{V_1 \over 2 }, \qquad
{d\xi_0  \over dt } = 2\mu _0,
\end{equation}
where $\mu _0 $ is the average velocity and $ \xi _{0} = {1 \over N }
\sum_{j=1}^{N} \xi _j$, is the center of mass of the $N $-soliton train.
The system of equations (\ref{eq:mu0-xi0}) for $V_2>0 $ has a simple
solution
\begin{equation}\label{eq:mu0-xi0s}
\mu _0(t) = \mu _{00} \cos (\Phi (t)), \qquad \xi _0(t) = \sqrt{ 2 \over
V_2 } \mu _{00} \sin (\Phi (t)) - {V_1  \over 2V_2 },
\end{equation}
where $\Phi (t)=\sqrt{2V_2}t +\Phi _0 $, and $\mu _{00} $ and $\Phi _0 $
are constants of integration. Therefore the overall effect of such
quadratic potential will be to induce a slow periodic motion of the train
as a whole.

Another important choice is the periodic potential
\begin{equation}\label{eq:V-per}
V^{(2)}(x) = A \cos(\Omega x+\Omega _0),
\end{equation}
where $A $, $\Omega  $ and $\Omega _0 $ are appropriately chosen
constants. NLS equation with similar potentials appear in a natural way in
the study of Bose-Einstein condensates, see \cite{BCCD}.

For $N=2 $ solitons the corresponding Karpman-Solov'ev system was
derived in \cite{UGL}. For $N>2 $ we obtain the PCTC where the integrals
for $N_k $, $M_k $, $\Xi_k $ and $D_k $ are equal to \cite{Liss}:
\begin{eqnarray}\label{eq:P16.1}
N_k^{(2)}&=&0, \qquad M_k^{(2)} = {\pi A\Omega^2 \over 8\nu _k }
{1 \over \sinh Z_k }\sin (\Omega \xi_k +\Omega _0), \\
\label{eq:P16.2}
\Xi_k^{(2)}&=&0, \qquad D_k^{(2)} =- {\pi^2 A\Omega ^2 \over 16\nu_k^2 }
{\cosh Z_k  \over \sinh ^2 Z_k } \cos (\Omega \xi_k +\Omega _0) ,
\end{eqnarray}
where $Z_k =\pi\Omega /(4\nu _k) $. These results allow one to derive the
corresponding perturbed CTC models. Again we find that $d\nu _0/dt =0 $.

\section{Analysis of the perturbed CTC and comparison with
         numerical simulations} \label{sec:Num}

The dynamics of an individual soliton in a train is determined by
the combined action of external potential and the influence of
neighboring solitons. The interaction with neighboring solitons
can be either repulsive, or attractive depending on the phase
relations between them. Particularly, if their amplitudes are
equal and the initial phase difference between neighboring
solitons is $\pi$ (as considered below) they repel each other
giving rise to expanding motion in the absence of an external
field \cite{GKUE,GUED}.

The external potential counterbalances the expansion, trying to
confine solitons in the minima of the potential. It is the
interplay of these two factors - the interaction of solitons and
the action of the external potential, which gives rise to a rich
dynamics of the $N$-soliton train.

To verify the adequacy of the perturbed CTC model for the description
of the $N$-soliton train dynamics in external potentials we
performed comparison of predictions of corresponding perturbed CTC
(PCTC) system and direct simulations of the underlying NLS
equation (\ref{eq:nls}). Below we present results pertaining to a
matter-wave soliton train in a confining (i) parabolic trap and in
(ii) a periodic potential modelling an optical lattice.

Here we present the numerical verification of the PCTC model.
The perturbed NLS eq. (\ref{eq:nls}) is solved by the operator splitting
procedure using the fast Fourier transform \cite{taha}. In the
course of time evolution we monitor the conservation of the norm and energy
of the N-soliton train. The corresponding PCTC equations are solved by the
Runge-Kutta scheme with the adaptive stepsize control \cite{numrecipes}.

The evolution of a $N$-soliton train in the absence of potential
($V(x)=0$) is well known, see e.g. \cite{GUED,GKDU}. These papers propose
a method to determine the asymptotic dynamical regime of the CTC for a
given set of initial parameters $\nu _k(0) $, $\mu _k(0) $, $\xi _k(0) $
and $\delta _k(0) $. Below we will use mainly the following set of
parameters:
\begin{eqnarray}\label{eq:in-par}
\nu _k(0) =1/2, \qquad \mu _k(0) =0, \qquad \xi_{k+1}(0) -\xi_k(0)= r_0,
\qquad  \delta _k(0) =k\pi,\\
\label{eq:in-para}
\nu _k(0) =1/2, \qquad \mu _k(0) =0, \qquad \xi_{k+1}(0) -\xi_k(0)= r_0,
\qquad  \delta _k(0) =0.
\end{eqnarray}
These two types of initial conditions (IC) are most widely used in numeric
simulations.

In the absence of potential the IC (\ref{eq:in-par}) ensure the so called
free  asymptotic regime, i.e.  each soliton develops its own velocity and
the distance between the neighboring solitons increases linearly in time.
At the same time the center of mass of the soliton train stays at rest
(see the left panel of Fig.~\ref{fig1}).  Under the IC (\ref{eq:in-para})
the solitons attract each other going into collisions whenever the
distance between them is not large enough.

From mathematical point of view the IC (\ref{eq:in-par}) reduce the CTC
into a standard (real) Toda chain for which the free asymptotic regime is
the only possible asymptotical regime. On the contrary the IC
(\ref{eq:in-para}) lead to singular solutions for the CTC (see
\cite{GKUE,GKDU,GEI}). The singularities of the exact solutions for the
CTC coincide with the positions of the collisions.

Below we will study the effects of the potentials for both types of IC.
One may expect that the quadratic potential will prevent free asymptotic
regime of IC (\ref{eq:in-par}) no matter how small $V_2>0 $ is and would
not be able to prevent collisions in the case of IC (\ref{eq:in-para}).
The periodic potential, if strong enough should be able to stabilize and
bring to bound states both types of IC.

\subsection{Quadratic potential}

For the quadratic external potential $V(x) = V_2 x^2 + V_1 x + V_0$ the
perturbed CTC equations in terms of soliton parameters have the form:

\begin{eqnarray}\label{eq:mu-k}
{d\mu _k \over dt } &=& 16\nu _0^3 \left( e^{-2\nu _0 (\xi_{k+1}
-\xi_{k})} \cos \left( 2\mu _0(\xi_{k+1} -\xi_{k}) +\delta _{k}
-\delta _{k+1} \right)\right.  \\ &-& \left. e^{-2\nu _0 (\xi_{k}
-\xi_{k-1})} \cos \left( 2\mu _0(\xi_{k} -\xi_{k-1}) +\delta
_{k-1} -\delta _{k} \right) \right) -V_2\xi_k - {V_1 \over 2 },
\nonumber \\
\label{eq:nu-k} {d\nu _k \over dt } &=& 16\nu _0^3 \left(
e^{-2\nu _0 (\xi_{k+1} -\xi_{k})} \sin \left( 2\mu _0(\xi_{k+1}
-\xi_{k}) +\delta _{k} -\delta _{k+1} \right)\right.  \\
&-& \left. e^{-2\nu _0 (\xi_{k} -\xi_{k-1})} \sin \left( 2\mu
_0(\xi_{k} -\xi_{k-1}) +\delta _{k-1} -\delta _{k} \right) \right),
\nonumber\\
\label{eq:xi-k}
{d\xi _k \over dt } &=& 2\mu _k, \\
\label{eq:del-k} {d\delta _k \over dt } &=& 2(\mu _k^2 +\nu _k^2)
+ V_2 \left( {\pi^2 \over 48\nu _k^2} -\xi_k^2\right) -V_1\xi_k
-V_0 .
\end{eqnarray}
where $\mu _k $, $\nu _k $, $\xi_k $ and $\delta _k $ for $k=1,\dots,N $
are the $4N $ soliton parameters, see eqs.
(\ref{eq:IC})-(\ref{eq:nls-1s}).

\begin{figure}[htb]
\centerline{\includegraphics[width=6cm,height=6cm,clip]{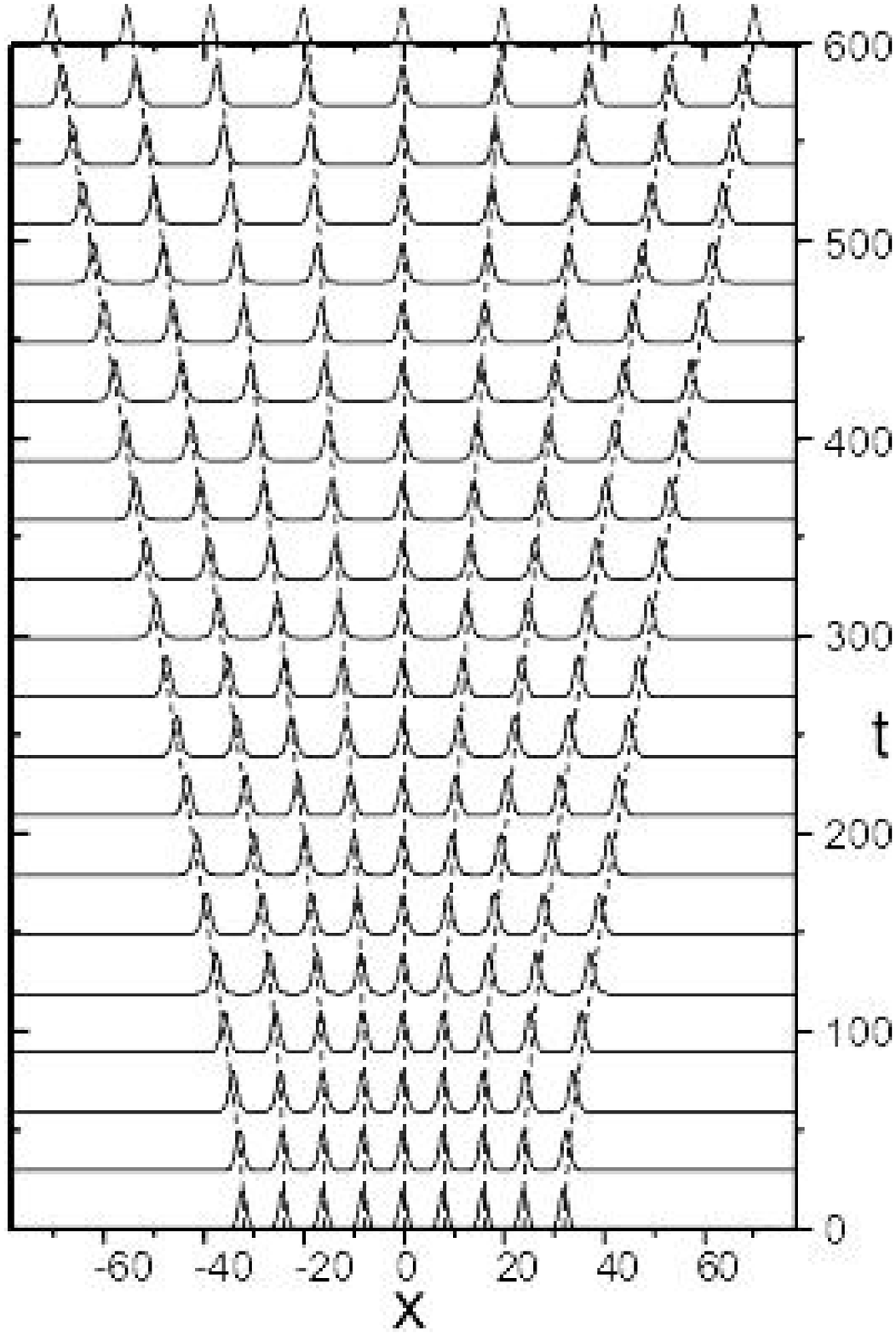}
            \includegraphics[width=6cm,height=6cm,clip]{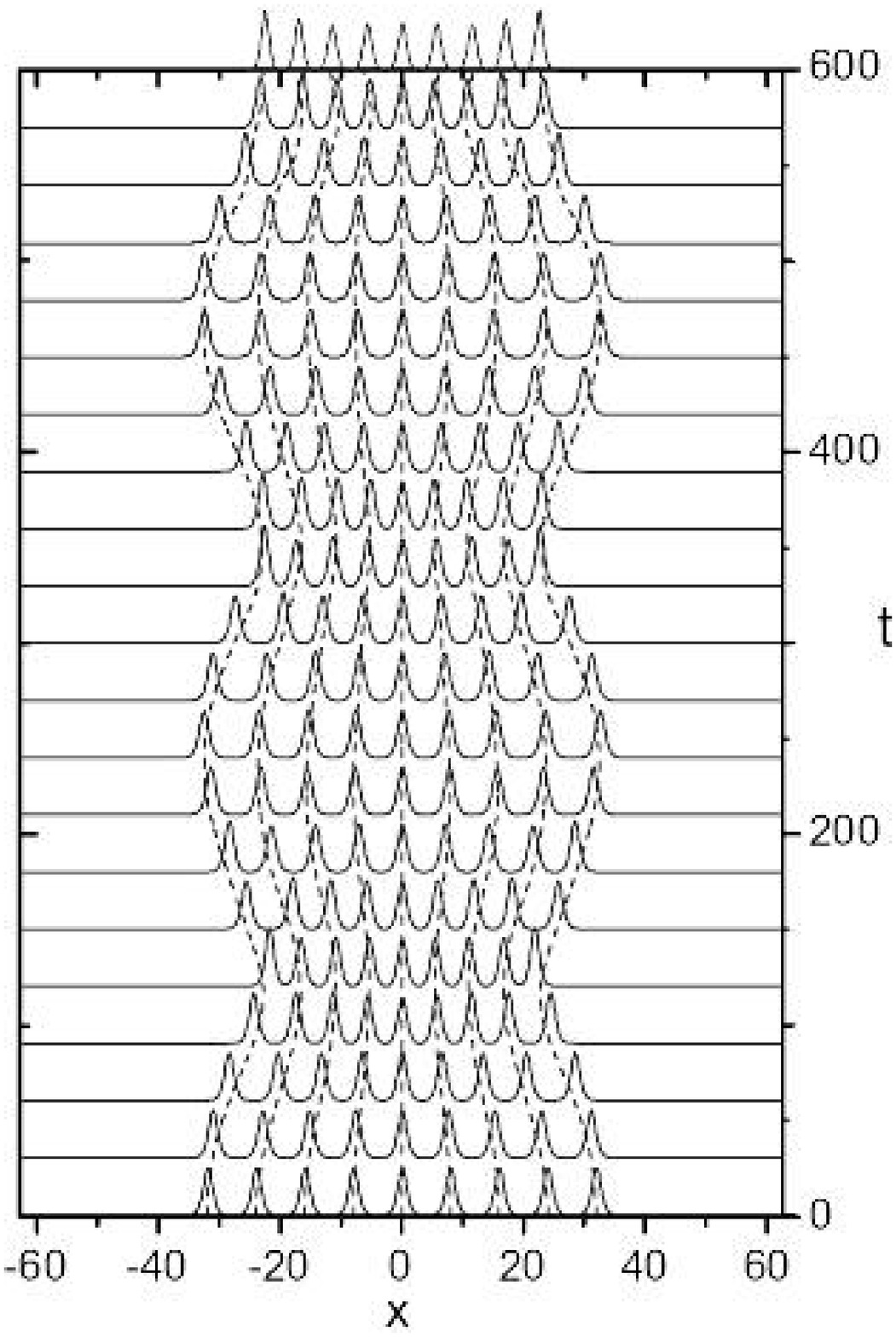}}
\caption{Left panel: $9$-soliton train with initial parameters as in
eq. (\ref{eq:in-par}) with $r_0=8 $ in the absence of a potential  goes into
free asymptotic regime.  Solid lines:  direct numerical simulation of the
NLS equation (\ref{eq:nls});  dashed lines:  $\xi_k(t) $ as predicted by
the CTC equations (\ref{eq:mu-k})-(\ref{eq:del-k}) with $V_0 = V_1 = V_2 =
0$ and $r_0=8 $.  Right panel:  Evolution of a $9$-soliton train with
the same initial parameters in the quadratic potential $V(x) = V_2x^2$
with $V_2=0.00005 $. Solid lines: direct numerical simulation  of the NLS
equation (\ref{eq:nls}); dashed lines: solution of the PCTC equations
(\ref{eq:mu-k})-(\ref{eq:del-k}).  Initially the train is placed
symmetrically relative to the minimum of the potential at $x=0$. }
\label{fig1}
\end{figure}

\begin{figure}[htb]
\centerline{\includegraphics[width=6cm,height=6cm,clip]{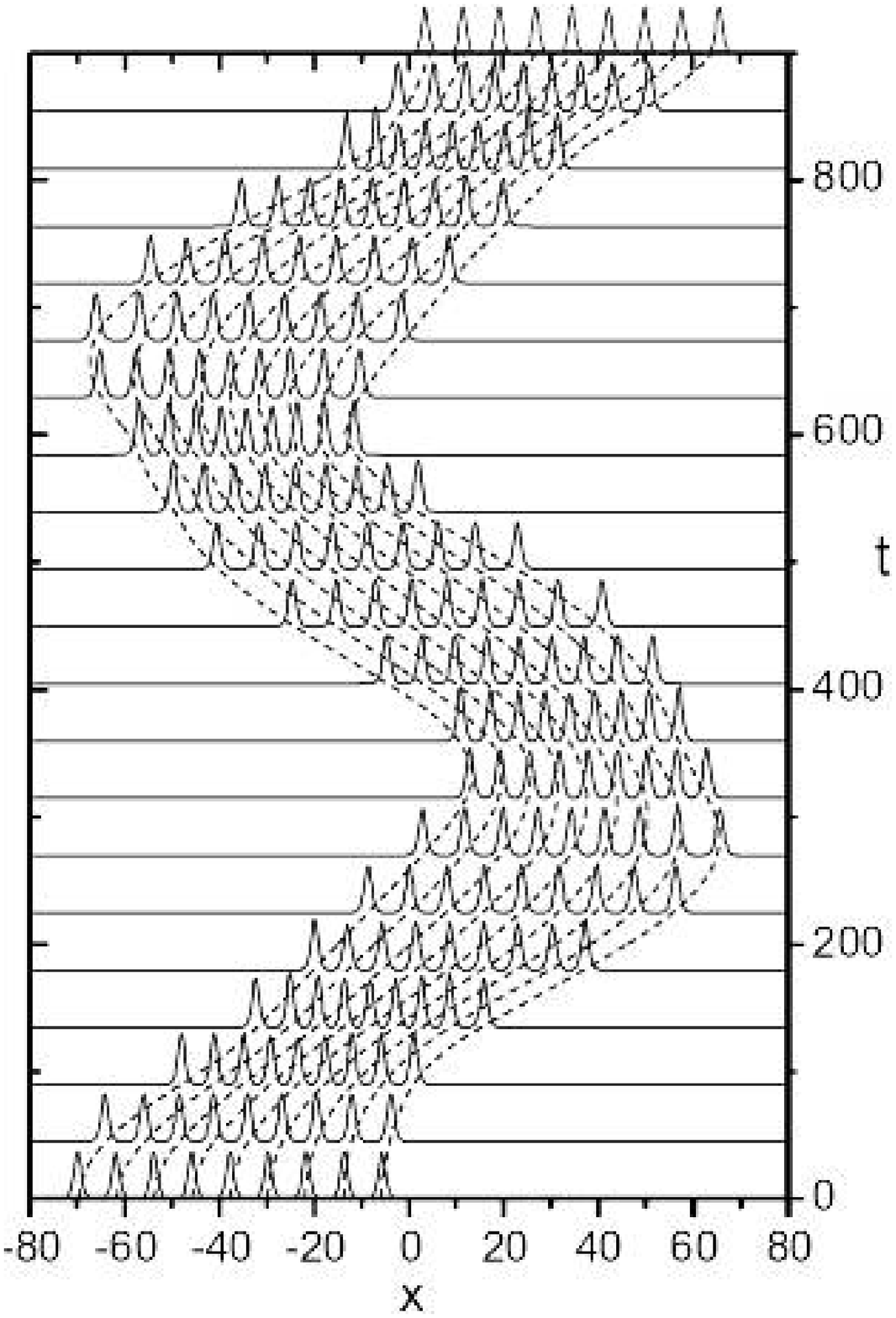} \qquad
            \includegraphics[width=6cm,height=6cm,clip]{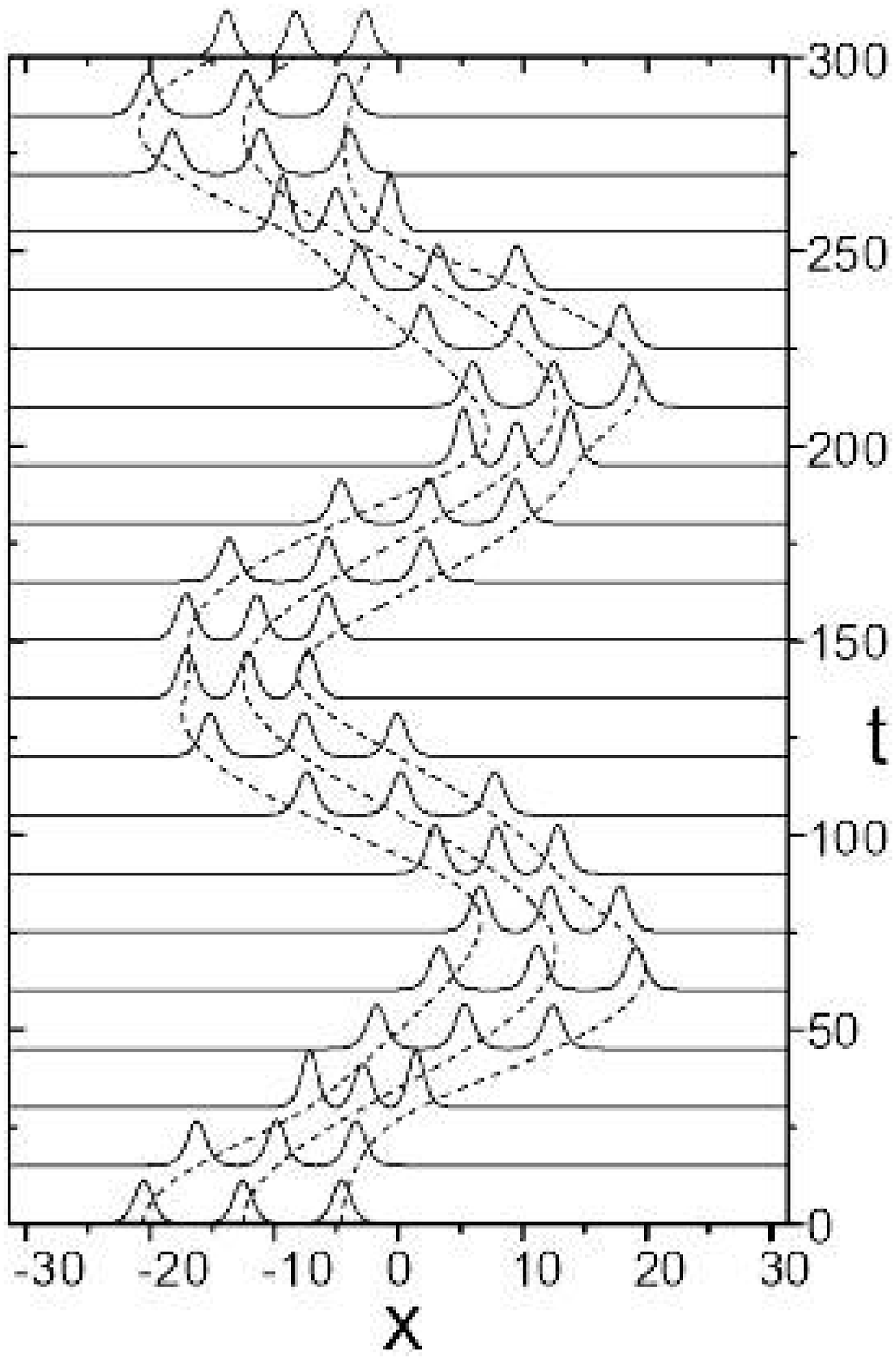}}
\caption{Harmonic oscillations of a N-soliton train initially shifted
relative to the minimum of the quadratic potential $V(x) = V_2x^2$.
Left panel: 9-soliton train, $V_2 = 0.00005$.
Right panel: 3-soliton train, $V_2 = 0.001$.
The IC of the both trains is given by (\ref{eq:in-par}) with $r_0=8 $.
In both panels solid lines correspond to direct simulations of the NLS
equation (\ref{eq:nls}), and dashed lines to numerical solution of the
PCTC equations (\ref{eq:mu-k}) - (\ref{eq:del-k}).}
\label{fig2}
\end{figure}

\begin{figure}[htb]
\centerline{\includegraphics[width=7cm,height=6cm,clip]{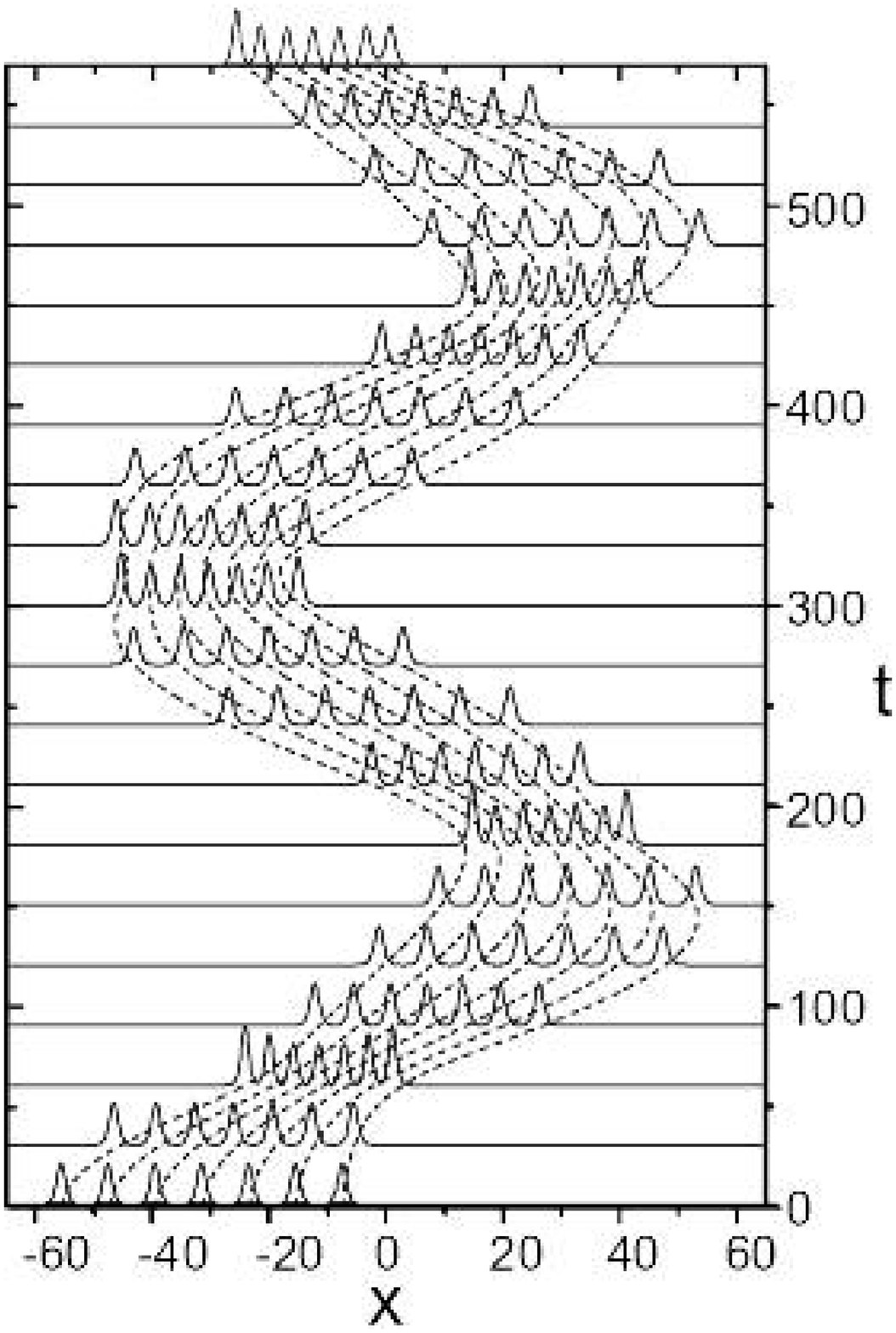}}
\caption{Dynamics of a 7-soliton train placed asymmetrically relative to the
minimum of the trap $V(x) = 0.0002 x^2$.
Solid lines: results of direct numerical simulations of the NLS
equation (\ref{eq:nls}). Dashed lines: result of solution of the PCTC
system (\ref{eq:mu-k}) - (\ref{eq:del-k}) for the center of mass
$\xi_i$. The parameters of solitons are the same as in (\ref{eq:in-par})
with $r_0=8$. The initial shift of the soliton train relative to the
minimum of the parabolic trap is $10 \pi$. }
\label{fig3}
\end{figure}

The effect of the quadratic potential on the $N $-soliton train with
parameters (\ref{eq:in-par}) is to balance the repulsive interaction
between the solitons, so that they remain bounded by the potential, as
illustrated in figures: the right panel of \ref{fig1}  and in \ref{fig2}.
The quadratic potentials are supposed to be weak, i.e. we choose $V_2 $ so
that
\begin{equation}\label{eq:v2}
V_2 \xi_N^2(0) \leq \nu _0, \qquad V_2 \xi_1^2(0) \leq \nu _0.
\end{equation}
Figures \ref{fig1} and \ref{fig2} show good agreement between the PCTC
model and the numerical solution of the perturbed NLS equation
(\ref{eq:nls}).  They also show two types of effects of the quadratic
potential on the motion of the $N $-soliton train: (i) the train performs
contracting and expanding oscillations if its center of mass coincides
with the minimum of the potential, (ii) the train oscillates around the
minimum of the potential as a whole if its center of mass is shifted. In
the last case contracting and expanding motions of the soliton train is
superimposed to the center of mass dynamics.  As one can see from the
figures the period of this motion matches very well the one predicted by
formula (\ref{eq:mu0-xi0s}). Indeed, from eq. (\ref{eq:mu0-xi0s}) it
follows that the period period of the center of mass motion is
$T=2\pi/\sqrt{2 V_2} $.  For the parameters in fig.~\ref{fig2} we have $T
\simeq 628$ (for 9-soliton train), $T \simeq 140$ (for $3$-soliton train).
Similar is the dynamics also for the $7 $-soliton train on
Fig.~\ref{fig3};  for the parameters choosen there we have $T=314.2$, in
good agreement with the numerical simulations.
The direct simulations of the NLS equation (\ref{eq:nls}) shows that
stronger parabolic trap may cause merging of individual solitons at times
of contraction, and restoring of the original configuration when the train
is expanded. This behavior reminds the phenomenon of "missing solitons"
observed in the experiment \cite{strecker}. However, this situation is
beyond the validity of the PCTC approach.

\subsection{Periodic potential}

Another external potential in which the $N$-soliton train exhibits
interesting dynamics is the periodic potential of the form $V(x) = A
\cos(\Omega x + \Omega_0)$. This case also may have a direct relevance to
matter - wave soliton trains confined to optical lattices.
The PCTC system in terms of soliton parameters has the form:
\begin{eqnarray}\label{eq:mu-k-p}
{d\mu _k \over dt } &=& 16\nu _0^3 \left( e^{-2\nu _0 (\xi_{k+1}
-\xi_{k})} \cos \left( 2\mu _0(\xi_{k+1} -\xi_{k}) +\delta _{k}
-\delta _{k+1} \right)\right. \nonumber \\ &-& \left. e^{-2\nu _0
(\xi_{k} -\xi_{k-1})} \cos \left( 2\mu _0(\xi_{k} -\xi_{k-1})
+\delta _{k-1} -\delta _{k} \right) \right) + M_k^{(2)}(\nu _k),\\
\label{eq:nu-k-p}
{d\nu _k \over dt } &=& 16\nu _0^3 \left(
e^{-2\nu _0 (\xi_{k+1} -\xi_{k})} \sin \left( 2\mu _0(\xi_{k+1}
-\xi_{k}) +\delta _{k} -\delta _{k+1} \right)\right.  \\
&-& \left. e^{-2\nu _0 (\xi_{k} -\xi_{k-1})} \sin \left( 2\mu
_0(\xi_{k} -\xi_{k-1}) +\delta _{k-1} -\delta _{k} \right) \right),
\nonumber\\
\label{eq:xi-k-p}
{d\xi _k \over dt } &=& 2\mu _k, \\
\label{eq:del-k-p} {d\delta _k \over dt } &=& 2(\mu _k^2 +\nu _k^2)
+ D_k^{(2)}(\nu _k),
\end{eqnarray}
where $ M_k^{(2)}(\nu _k)$ and $D_k^{(2)}(\nu _k)$ are given in
(\ref{eq:P16.1}) and (\ref{eq:P16.2}).

\begin{figure}[htb]
\centerline{\includegraphics[width=7cm,height=5cm,clip]{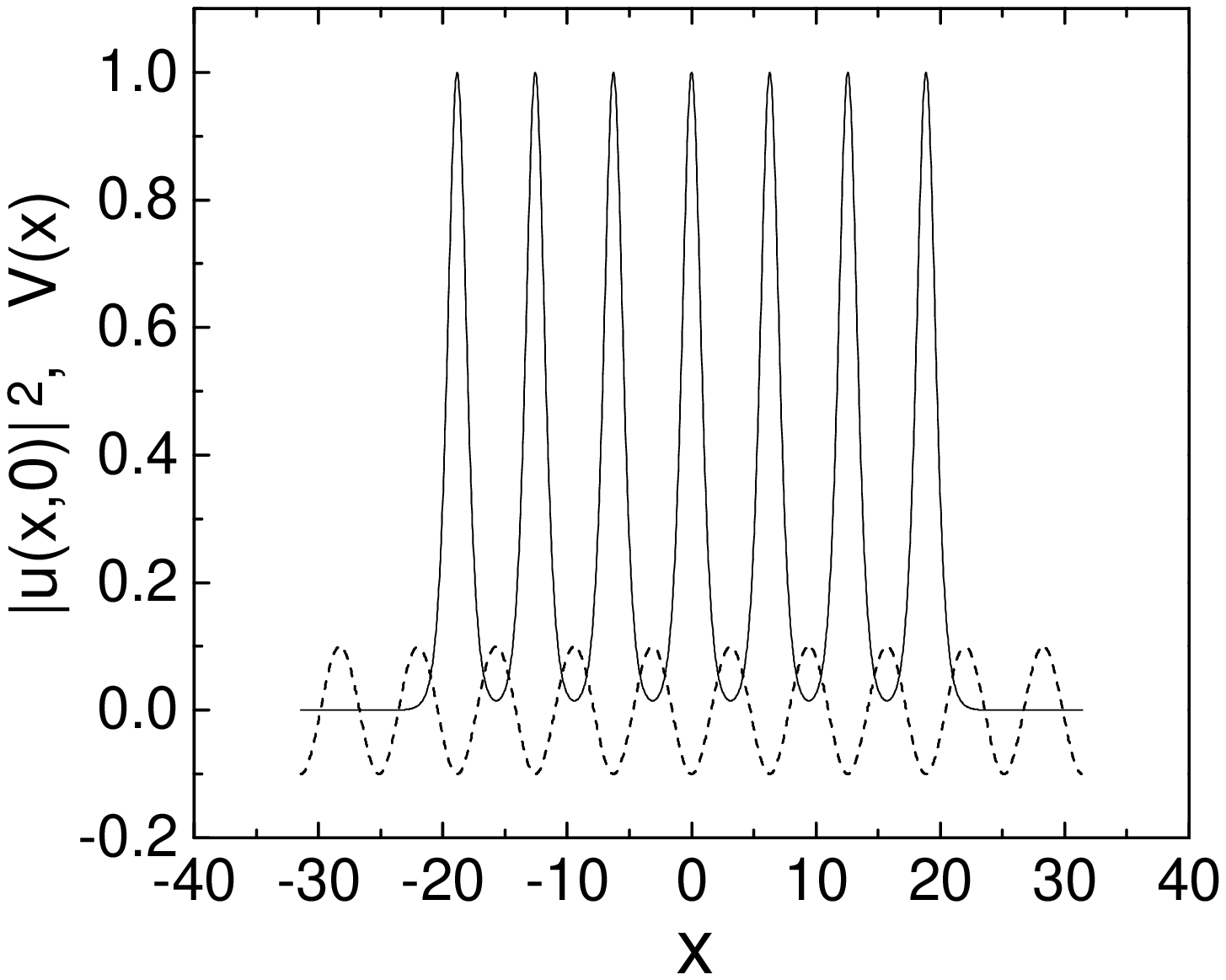}}
\caption{Solitons (continuous line) remain confined around the
minima of the periodic potential $V(x) = A \cos(x)$ (dashed line)
performing small amplitude oscillations if its strength is big
enough $A=0.1$.}
\label{fig4}
\end{figure}

\begin{figure}[ht]
\centerline{\includegraphics[width=6cm,height=6cm,clip]{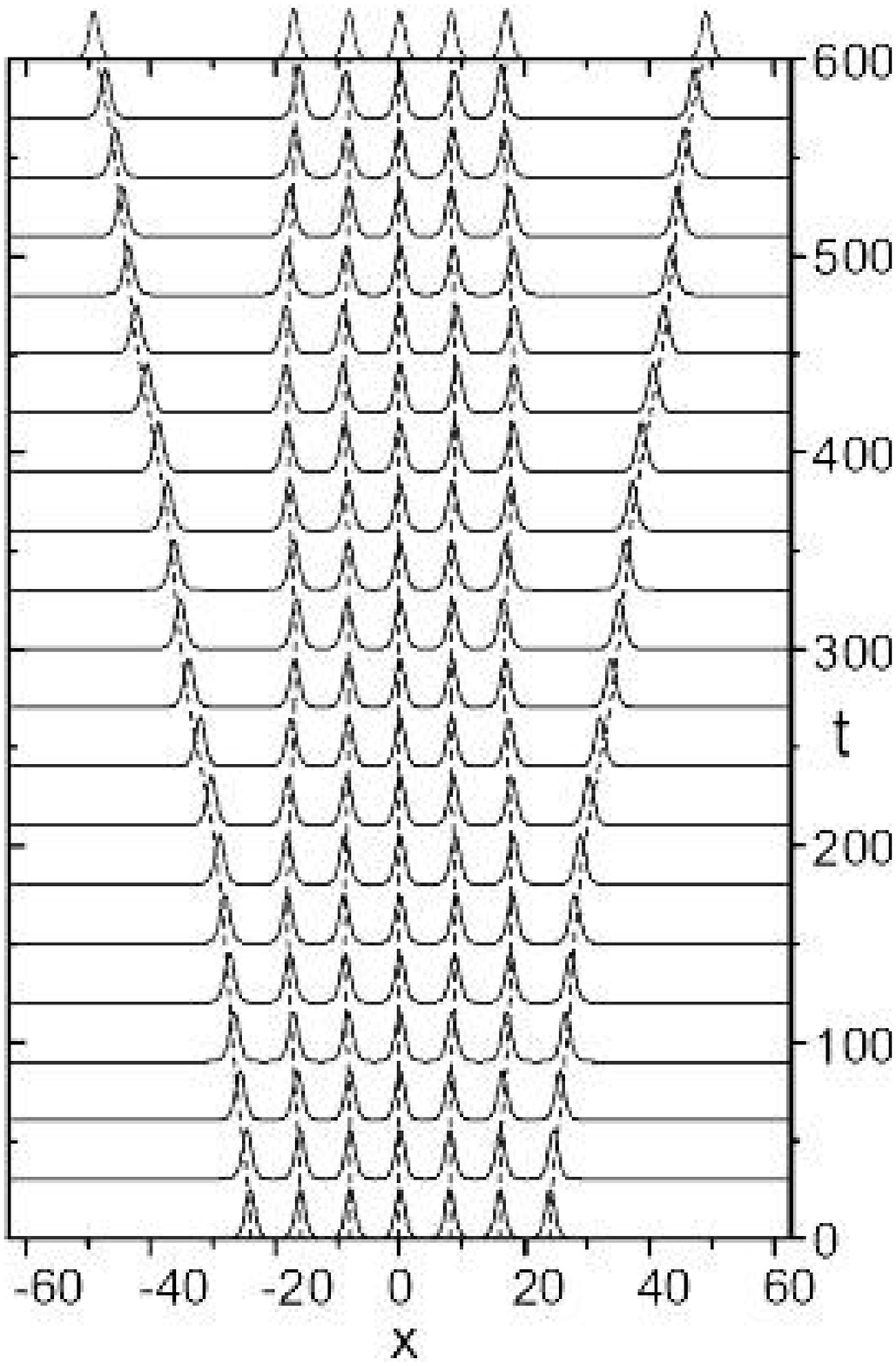} \qquad
            \includegraphics[width=6cm,height=6cm,clip]{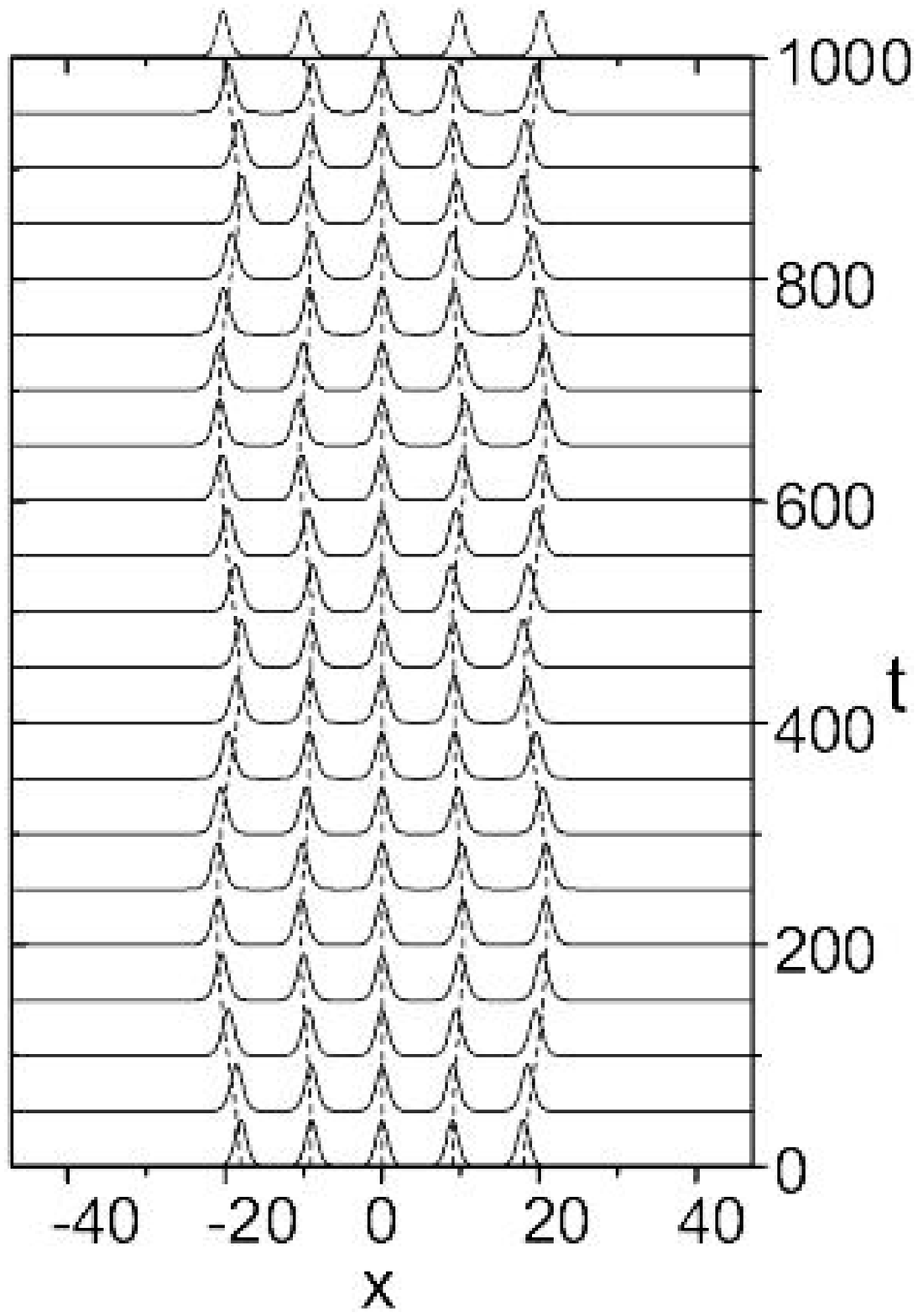}}
\caption{Left panel: The expulsion of solitons from the train, as obtained
from direct simulations of the NLS equation (\ref{eq:nls}) (solid lines),
and as predicted by PCTC system (\ref{eq:mu-k-p})-(\ref{eq:del-k-p}) for the
center of mass $\xi_i$ (dashed lines).  The IC of the $7 $-soliton train
are given by (\ref{eq:in-par}) with $r_0=8 $;
the parameters of the periodic potential $V(x) = A
\cos(\Omega x + \Omega_0)$ are $A= 0.001, \ \Omega = \pi/4, \ \Omega_0 =
0$. Right panel: Oscillations of the 5-soliton train
with IC given by (\ref{eq:in-par}) with $r_0=9 $; in a moderately weak
periodic potential, $A = 0.0005$, $\Omega = 2\pi/9$, $\Omega _0 = 0$.
Solid and dashed lines correspond, respectively, to numerical solution of
the NLS eq. (\ref{eq:nls}) and PCTC system
(\ref{eq:mu-k-p}) - (\ref{eq:del-k-p}). }
\label{fig5}
\end{figure}

Each soliton of the train experience confining force of the periodic
potential and repulsive force of neighboring solitons. Therefore,
equilibrium positions of solitons do not coincide with the minima of the
periodic potential. Solitons placed initially at minima of the periodic
potential (Fig. \ref{fig4}) perform small amplitude oscillations around
these minima, provided that the strength of the potential is big enough to
keep solitons confined. As opposed, the weak periodic potential is unable
to confine solitons, and repulsive forces between neighboring solitons (at
phase difference $\pi$) induces unbounded expansion of the train.
In the intermediate region, when the confining force of the
periodic potential is comparable with the repulsive forces of
neighboring solitons, interesting dynamics can be observed such as
the expulsion of bordering solitons from the train, as shown in the left
panel of Fig. \ref{fig5}.
This phenomenon, revealing the complexity of the internal dynamics
of the train,  can be explained as follows. Each soliton performs
nonlinear oscillations within individual potential wells under
repulsive forces from neighboring solitons. When the amplitude of
oscillations of particular solitons grow and two solitons closely
approach each other, a strong recoil momentum can cause the
soliton to leave the train, overcoming barriers of the periodic
potential. In Fig. \ref{fig5} this happens with bordering solitons
(the other  solitons remain bounded under long time evolution). It
is noteworthy to stress that this phenomenon is well described by
the PCTC model, as is evident from Fig. \ref{fig5}, left panel.

On the right panel of the same figure we have similar IC
as in (\ref{eq:in-par}) and we have choosen again
the initial positions of the solitons to coincide with the minima
of the periodic potential $V(x) = A \cos(\Omega x +\Omega_0)$; i.e.
$r_0=2\pi/\Omega  $. The values of $A=0.0005 $ and $r_0=9 $ in the right
panel of Fig. \ref{fig5} now are such that the solitons form a bound
state. Therefore for any given initial distance $r_0 $ there is a critical
value  $A_{\rm cr}(r_0) $ for $A $ such that for $A>A_{\rm cr}(r_0) $ the
soliton train with IC (\ref{eq:in-par}) will form a bound state.

In contrast to the quadratic potentials, the weak periodic potential is
unable to confine solitons, and repulsive forces between neighboring
solitons (at $\nu _k(0) =1/2, \ \delta _k(0) =k\pi$) induces unbounded
expansion of the train similar to what was shown in the left panel of
Fig. \ref{fig1}.

\begin{figure}[ht]
\centerline{\includegraphics[width=6cm,height=6cm,clip]{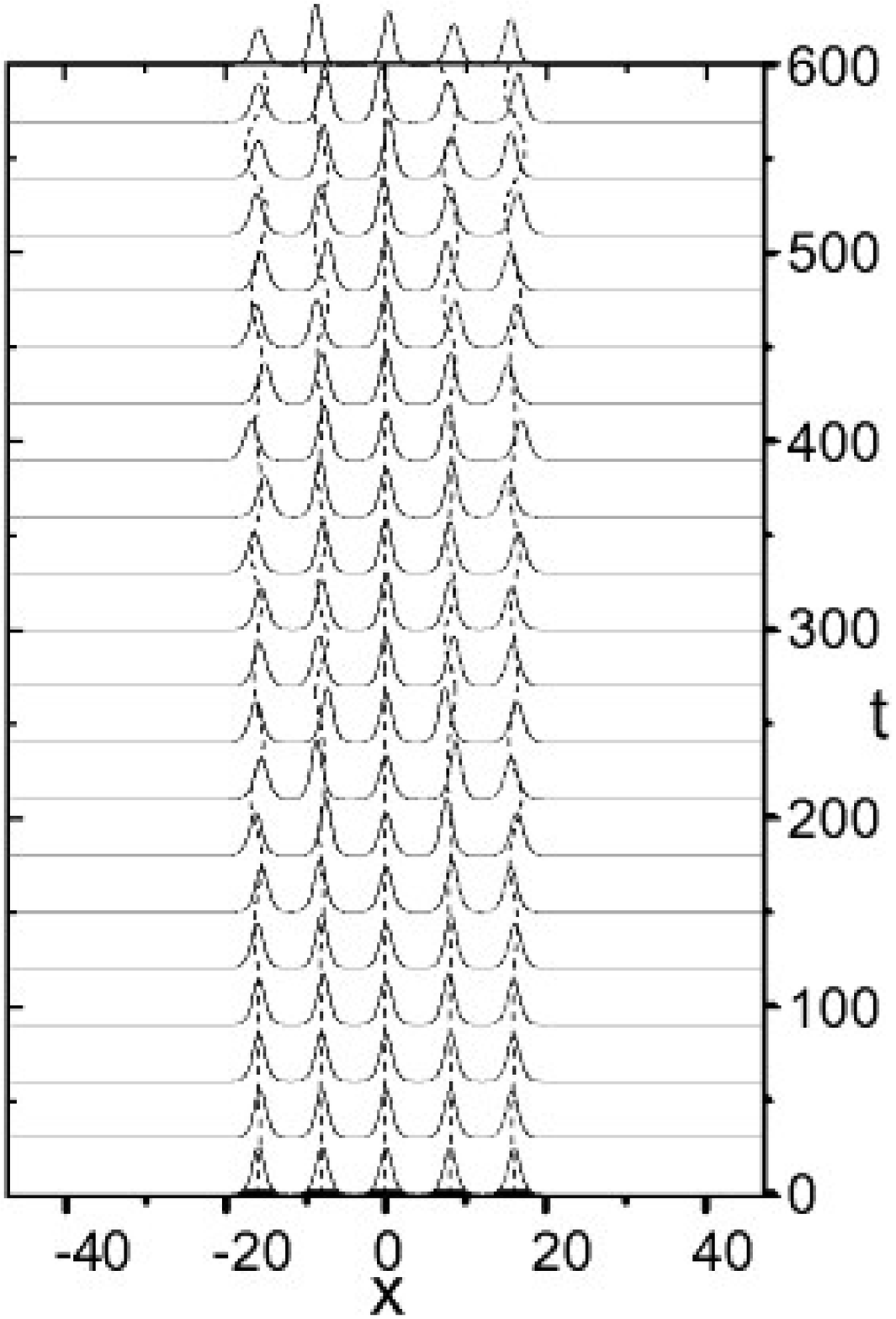} \qquad
            \includegraphics[width=6cm,height=6cm,clip]{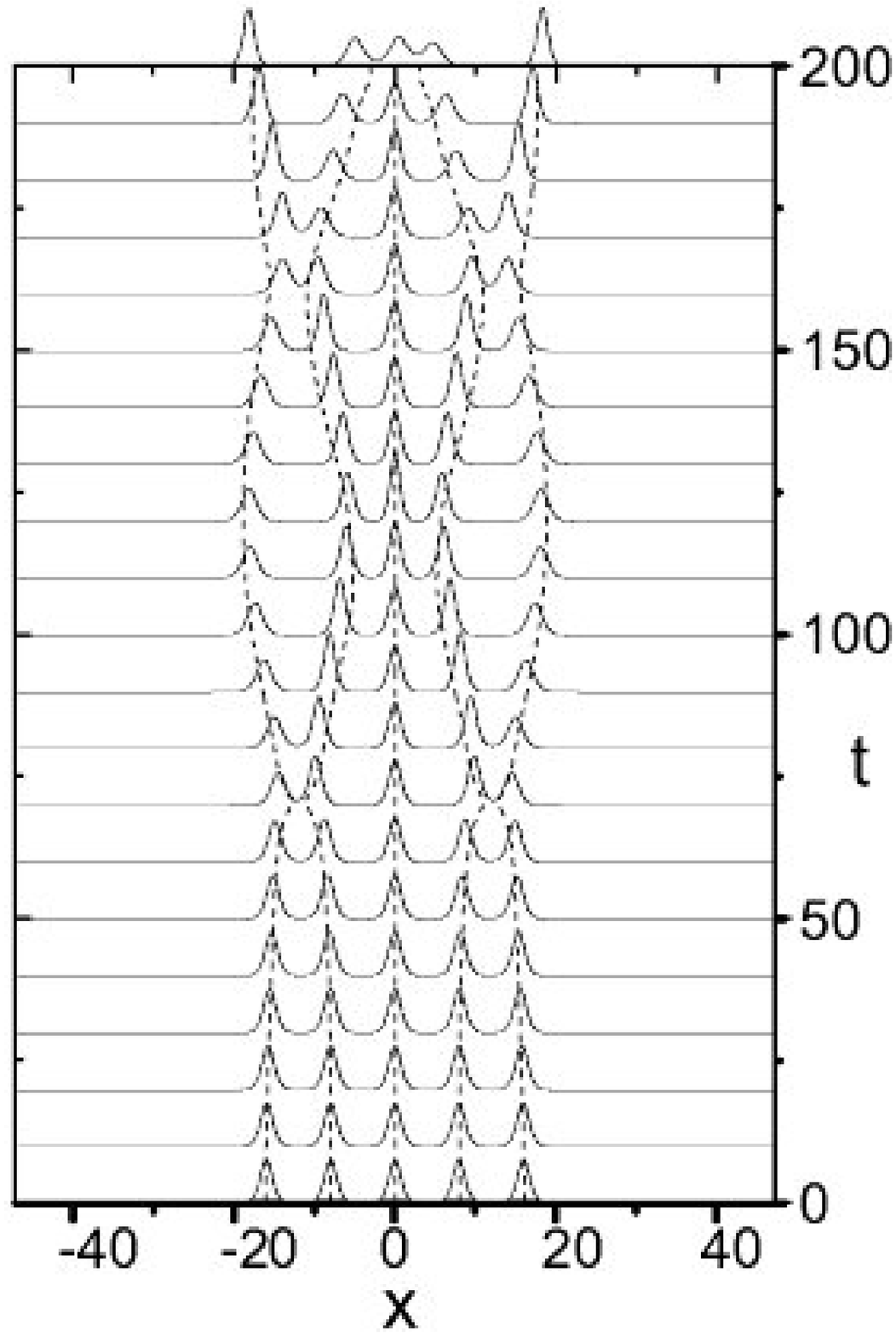}}
\caption{Dynamics of a 5-soliton train with zero phase difference
between neighboring solitons, in the periodic potential
$V(x) = A \cos(\Omega x +\Omega_0)$ with $\Omega = \pi/4$, $\Omega_0=0$,
and $r_0 = 8$.
Left panel: When the periodic potential is strong enough
$A = 0.02$, the N-soliton train remain confined, each soliton performing
small amplitude oscillations around the minima of individual cells.
Right panel: Weaker periodic potential $A=0.01$ cannot prevent solitons
from collisions, which destroy the train.
Solid and dashed lines correspond, respectively, to numerical solution of
the NLS eq. (\ref{eq:nls}) and PCTC system
(\ref{eq:mu-k-p}) - (\ref{eq:del-k-p}). }
\label{fig6}
\end{figure}

The periodic potential can play stabilizing role also for the IC
(\ref{eq:in-para}), when the zero phase difference between
neighboring solitons correspond to their mutual attraction.
If the periodic potential is strong enough, solitons do not experience
collision. The weak periodic potential cannot prevent solitons from
collisions, which eventually leads to destruction of the soliton train, as
illustrated in Fig. \ref{fig6}.
Again for any given initial distance $r_0 $ there will be a critical
value $A_{\rm cr}'(r_0) $ for $A $ such that for $A>A'_{\rm cr}(r_0) $ the
soliton train with IC (\ref{eq:in-para}) will form a bound state avoiding
collisions.

\subsection{Tilted periodic potential}

Now we consider the dynamics of a N - soliton train in a tilted periodic
potential, which is the combination of periodic and linear potentials
\begin{equation} \label{tilted}
V(x) = A \cos(\Omega x + \Omega_0) + B x.
\end{equation}
This potential is of particular interest in studies of
Bose-Einstein condensates. A train of repulsive BEC loaded in such
a potential (where the periodic potential was a 1D optical lattice
and the linear one was due to the gravitation) exhibited Bloch
oscillations \cite{anderson}. At each period of these oscillations
condensate atoms residing in individual optical lattice cells
coherently tunneled through the potential barriers. This was the
first experimental demonstration of a pulsed atomic laser
\cite{anderson}. Recently a new model of a pulsed atomic laser was
theoretically developed in \cite{carr}, where the solitons of
attractive BEC were considered as carriers of coherent atomic
pulses.

\begin{figure}[htb]
\centerline{\includegraphics[width=6cm,height=6cm,clip]{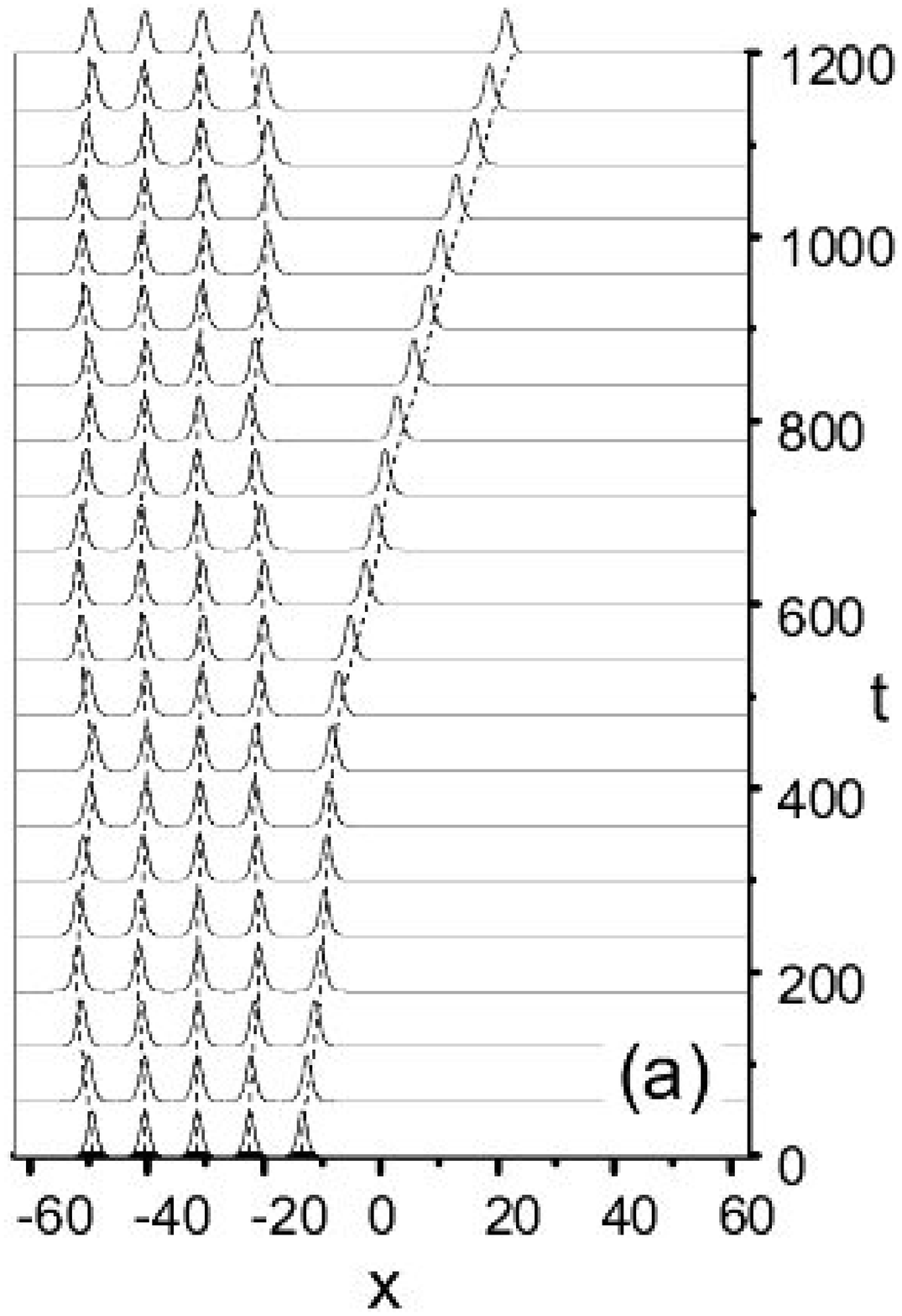}
            \includegraphics[width=6cm,height=6cm,clip]{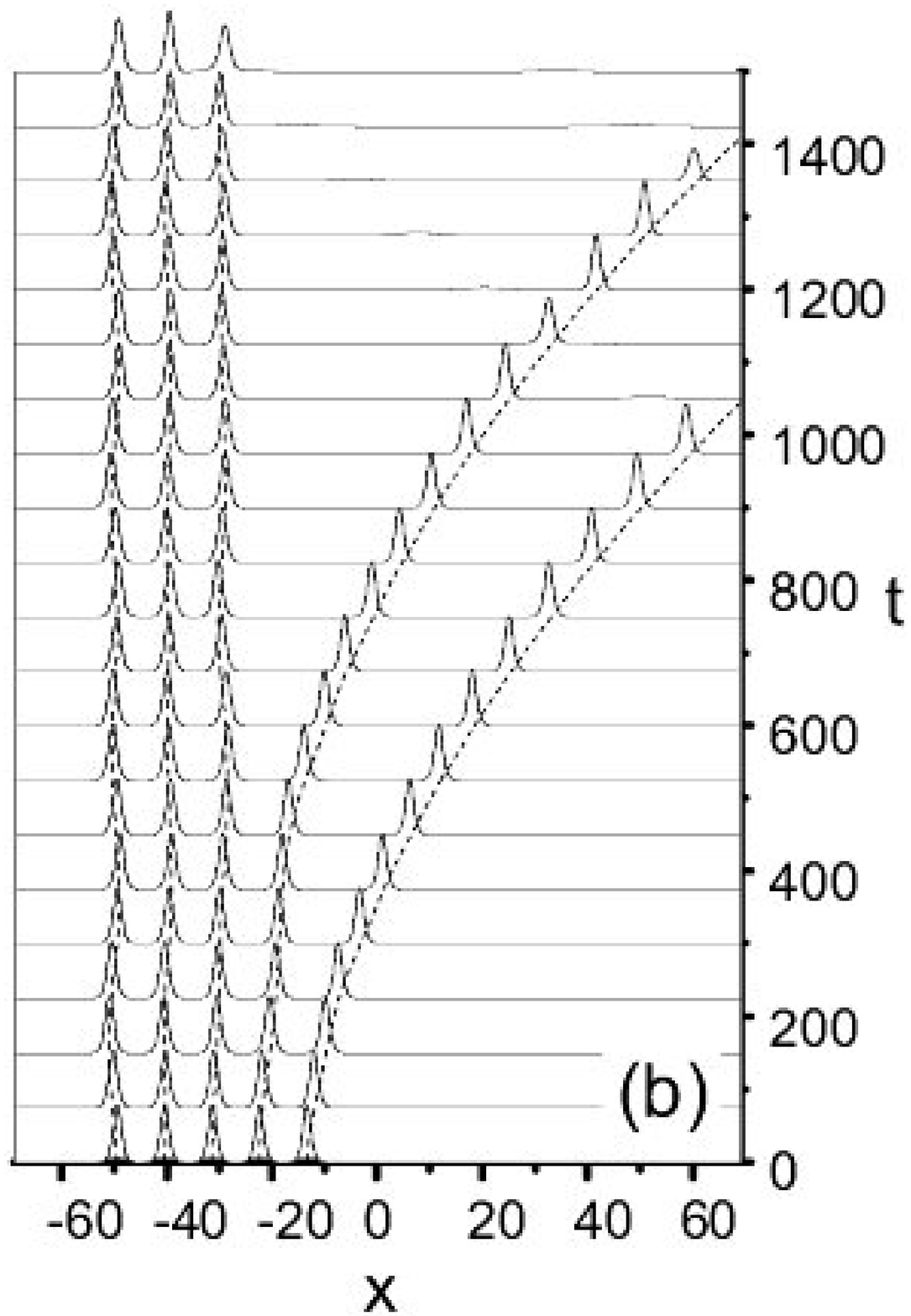}}
\centerline{\includegraphics[width=6cm,height=6cm,clip]{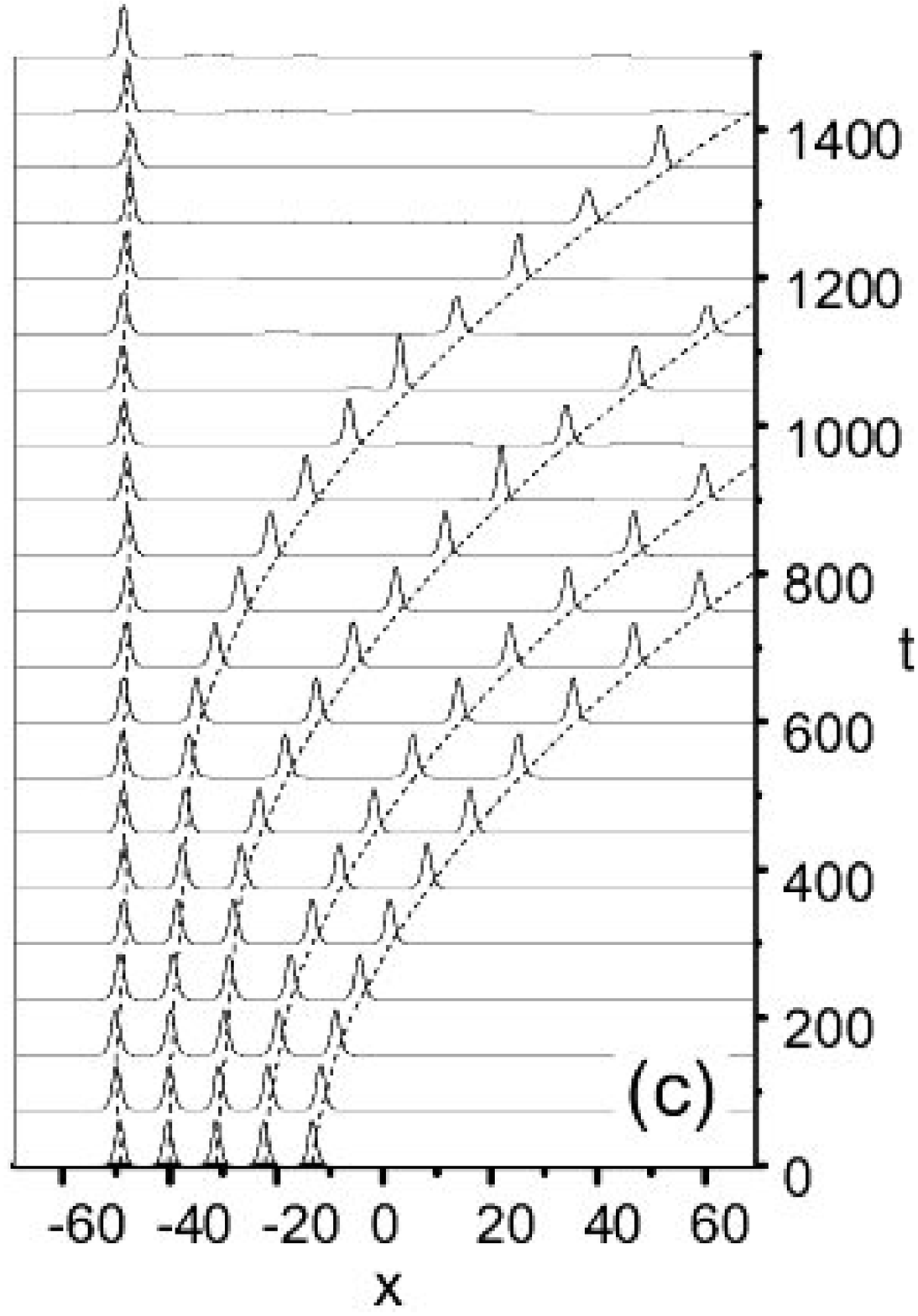}
            \includegraphics[width=6cm,height=6cm,clip]{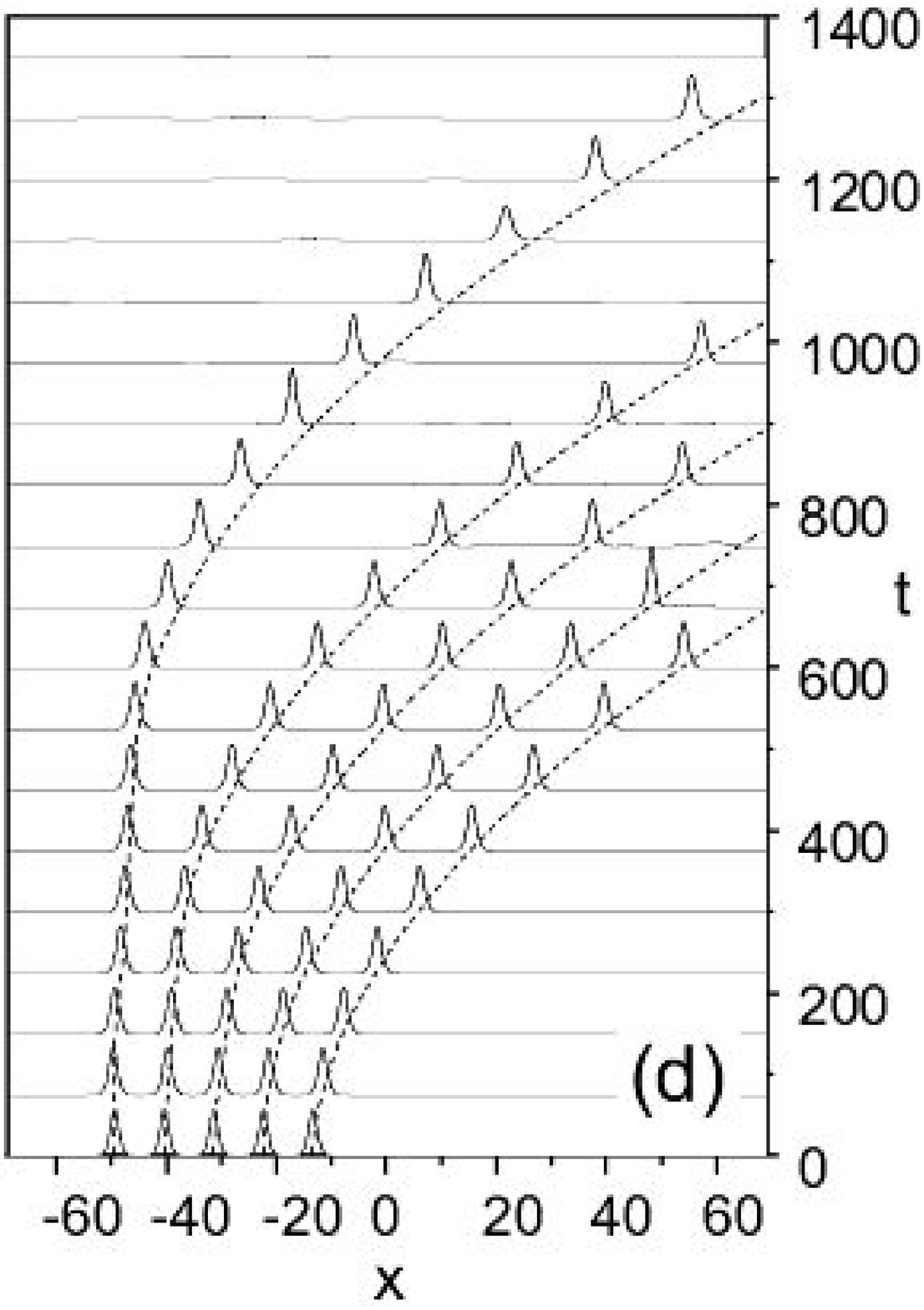}}
\caption{Controlled withdrawal of solitons from the 5-soliton
train by adjusting the strength of the linear potential
(\ref{tilted}) with parameters: $A = 0.0005$, $\Omega =2\pi/9$,
$\Omega_0 = 0$. Depending on the tilt, different number of
solitons can be pulled out of the train: (a) one soliton at $B = -
0.00003$, (b) two at $B = -0.00011$, (c) four at $B = -0.0002$ and
(d) five at $B = -0.0003$. The initial phase difference and
separation between neighboring solitons in the train are,
respectively, $\pi$ and $9$. Initially the train is shifted by
$-10 \pi$ with respect to $x=0$ for graphical convenience. Solid
and dashed lines correspond, respectively, to direct simulations
of the NLS equation (\ref{eq:nls}) and numerical integration of the PCTC
system (\ref{eq:mu-k-p}) - (\ref{eq:del-k-p}) .}
\label{fig7}
\end{figure}

Controlled manipulation with matter - wave solitons is important
issue in these applications. Below we demonstrate that solitons of
attractive BEC confined in tilted optical lattice can be flexibly
manipulated by adjustment of the strength of the linear potential.
In Fig. \ref{fig7} we show the extraction of different number of
solitons from the 5-soliton train by increasing the strength of
the linear potential $B$, as obtained from direct simulations of
the NLS equation (1) and numerical integration of the PCTC system
(38) - (41) with
\begin{eqnarray}
M_k^{(2)} &=& {\pi A\Omega^2 \over 8\nu _k }
{1 \over \sinh Z_k }\sin (\Omega \xi_k +\Omega _0) - \frac{1}{2} B, \label{mk}\\
D_k^{(2)} &=&- {\pi^2 A\Omega ^2 \over 16\nu_k^2 }
{\cosh Z_k  \over \sinh ^2 Z_k } \cos (\Omega \xi_k +\Omega _0)
-B \xi_k. \label{dk} \\
\end{eqnarray}
As is evident from Fig. \ref{fig7}, the PCTC model provides adequate
description of the dynamics of a N-soliton train in a tilted periodic
potential. A small divergence between predictions for the trajectory
of the left border soliton in Fig. \ref{fig7} (d), is due to the
imperfect absorption of solitons from the right end of the integration
domain. Reflected waves enter the integration domain and interact with
solitons, which causes the discrepancy.

\section{Conclusions} \label{sec:PCTC}

We have studied the dynamics of the $N$-soliton train confined to external
fields (quadratic and periodic potentials). Both the analytical treatment
in the framework of the PCTC model, and numerical analysis by direct
simulations of the underlying NLS equation show that the PCTC is adequate
for description of the $N$-soliton interactions in external potentials.
Relevance of this study to the research on matter-wave soliton trains
in magnetic traps and optical lattices is briefly mentioned.

Like any other model, the predictions of the CTC should be compared with
the numerical solutions of the corresponding NLEE. Such comparison between
the CTC and the NLS has been done thoroughly in \cite{GKUE,GUED,GKDU} and
excellent match has been found for all dynamical regimes. This means that
the CTC may be viewed as an {\em universal model} for the adiabatic
$N$-soliton interactions for several types of NLS.

For the perturbed CTC equations such comparison has been just started; the
good agreement shown in the figures above supports the hope that the region
of applicability of PCTC can be widened.

More detailed investigation of the $N $-soliton train interactions under
different types of external potentials and for diffetent types of
initial soliton parameters will be published in subsequent papers.

\section*{Acknowledgments}\label{sec:Ack}

V. S. G. is grateful to Professors M. Boiti, F. Pempinelli and B.
Prinari for giving the chance to participate in the Conference
"Nonlinear Physics, Theory and Experiment III" and for warm
hospitality at the University of Lecce, where part of this work
was done. Partial support from the Bulgarian Science
Foundation through contract No. F-1410 is acknowledged.
B. B. B. thanks the Department of Physics at the
University of Salerno, Italy, for a research grant. M. S.
acknowledges partial financial support from the MIUR, through the
inter-university project PRIN-2003, and from the Istituto
Nazionale di Fisica Nucleare, sezione di Salerno. We are grateful
to Prof. I. Uzunov for useful discussions and for calling our
attention to Refs. \cite{WB,UGL}.

\end{document}